\documentclass{PoS}

\usepackage{mathtools}
\usepackage{tikz-cd}
\usepackage{amsmath}
\usepackage{hhline}
\usepackage[amsmath]{empheq}
\usepackage{cancel}

\usepackage{mathrsfs}
\usepackage{tikz}

\newcommand{\dd}{\mathrm{d}}

\newcommand{\e}{{\epsilon}}
\newcommand{\w}{\wedge}
\newcommand{\bbm}{\left(\begin{matrix}}
\newcommand{\ebm}{\end{matrix}\right)}
\newcommand{\beq}{\begin{eqnarray}}
\newcommand{\eeq}{\end{eqnarray}}

\newcommand{\sfrac}[2]{{\textstyle\frac{#1}{#2}}}

\newcommand{\be}{\begin{equation}}
\newcommand{\ee}{\end{equation}}

\newcommand{\beqa}{\begin{eqnarray}}
\newcommand{\eeqa}{\end{eqnarray}} 
\def\nn{\nonumber} \def \bea{\begin{eqnarray}} \def\eea{\end{eqnarray}}

\newcommand{\barr}{\begin{array}}
\newcommand{\earr}{\end{array}}

 \def\d{\delta} 

    \def\r{\rho}
 \def\S{\Sigma}  

   \def\X{\mathbb X} \def \A{\mathbb A}

 \def\one{\mbox{1 \kern-.59em {\rm l}}}

\def\bit{\begin{itemize}} \def\eit{\end{itemize}}

\def\({\left(} \def\){\right)}

\title{$L_\infty$-algebras and membrane sigma models}

\ShortTitle{$L_\infty$-algebras and membrane sigma models}

\author{Clay J. Grewcoe, \speaker{Larisa Jonke}
\\
       Division of Theoretical Physics,        Rudjer Bo\v skovi\' c Institute, \\
       Bijeni\v cka cesta 54,  Zagreb, Croatia\\
        E-mail:  \email{cgrewcoe@irb.hr}, \email{larisa@irb.hr}}

\abstract{ Membrane sigma-models have been used for the systematic description of closed strings in non-geometric flux backgrounds.  In particular, the conditions for gauge invariance of the corresponding action functionals were related to the Bianchi identities for the fluxes. In this contribution we demonstrate how to express these Bianchi identities in terms of homotopy relations of the  underlying $L_\infty$-algebra for the case of the Courant sigma-model. 
We argue that this result can be utilized in understanding the constraint structure of Double Field Theory and the corresponding membrane sigma-model.

         }

\FullConference{Corfu Summer Institute 2019 "School and Workshops on Elementary Particle Physics and Gravity" (CORFU2019)\\
		31 August - 25 September 2019\\
		Corfu,  Greece}

\begin{document}

\section{Motivation and introduction}

Our understanding of physics in quantum gravity regime is far from complete, but there are certain insights providing basis for further advances.   In particular,   to resolve the geometry of space-time in that regime one should use extended probes instead of point particles. This also implies that the concept of symmetries should be generalized, and more often than not, that the field  content needs to be extended. In various  concrete settings   these  observations were made precise which lead to many interesting developments in   e.g. string theory, matrix/tensor models, and  higher-spin (gravity) theories. In this contribution we would like to review part of the  recent progress in the understanding of generalized symmetries relevant in Double Field Theory (DFT).

DFT is a proposal to incorporate T-duality  symmetry of string theory as a symmetry of a field theory defined on the double configuration space \cite{S1,S2,HZ1}.  Strings being  extended objects can, unlike point particles, wind around non-contractible cycles. When compactified on toroidal backgrounds, string theory enjoys T-duality symmetry under the  interchange of momentum and winding modes.  String backgrounds that are
T-dual to each other may correspond to target spaces with different geometry and topology. Moreover, a duality transformation can result in  unconventional closed string geometries where string
duality transformations are required as transition functions, and lead to non-geometric flux
backgrounds, see Ref.\cite{erik} for a review and references therein.  In order to account for this duality symmetry DFT is formulated on a configuration space extended by a  set of dual coordinates conjugate to winding modes of a string, and it  provided a natural setting for the description of non-geometric string backgrounds.  However, the theory is constrained in the sense that, although all coordinates are doubled, the physical fields and parameters depend only on half of them. The strong constraint\footnote{The weak constrained of DFT corresponds to level-matching condition in string theory and presents no additional constraint of the theory.} makes DFT a well-defined theory, but limits its physical applications as we shall argue latter on.  

The symmetries of DFT correspond to generalized diffeomorphisms, which combine standard diffeomorphisms with gauge transformations of a 2-form Kalb-Ramond field. These can naturally be described in the framework of generalized geometry defined on a doubled bundle \cite{HZ2}. Taking locally $TM\oplus T^\ast M$ as a doubled bundle over a target space manifold $M$ one can define  a skew-symmetric  bracket on its sections known as Courant bracket, symmetric pairing that corresponds to the $O(d,d)$ metric and a map from the bundle to $TM$.  These data, accompanied by a number of compatibility conditions,  define a Courant algebroid (CA) \cite{C90,LWX,Pavol1}. On the solution of the strong constraint the relevant bracket of the symmetry algebra of DFT reduces to the Courant bracket. Furthermore, one can twist the Courant bracket by a generalized 3-form which can be related to (non-geometric)  fluxes in DFT \cite{Blumenhagen}. A particularly nice way to describe these fluxes  in a systematic manner  is by using the  membrane sigma model \cite{Halmagyi, Mylonas, ChJL, Watamura}.  The starting  point is a three dimensional topological model known as the Courant sigma model, which can be uniquely defined using CA data. One can show that the conditions for gauge invariance of this sigma model produce relations for background fluxes and their Bianchi identities, again on a given solution of the strong constraint. 
A membrane sigma model for DFT was proposed in Ref.\cite{p1}, where it was  demonstrated
how the standard T-duality orbit of geometric and non-geometric flux backgrounds is captured by its action functional in a unified way. However, this membrane sigma model is gauge  invariant only after imposing the strong constraint. 

We propose that in order to obtain a better insight into the interplay between the gauge symmetry of sigma models, T-duality symmetry and the strong constraint we analyse the aforementioned sigma models  in $L_\infty$-algebra framework. $L_\infty$-algebras are generalizations of   Lie algebras with infinitely-many higher  brackets, related to each other by higher homotopy versions of the Jacobi identity \cite{Barton, Jim}. There are two aspects of an $L_\infty$-algebra structure that are relevant here. First, one can think of an $L_\infty$-algebra as a geometric structure corresponding to the BV-BRST complex. In that respect it is not surprising that one can use this framework to construct consistent deformations of a given (gauge-invariant, perturbative) theory, see e.g.\cite{bb}. Moreover, as one can incorporate both symmetries and dynamics in an $L_\infty$ structure, even for non-Lagrangian theories, one hopes to clearly identify the role of constraints.

 The first step toward this goal is to recast the Courant sigma model in terms of an $L_\infty$-algebra. This was done in Ref.\cite{p2} and here we shall review parts of this construction. In the next section we shall remind the reader of the rich  gauge structure of the  Courant sigma model. After a short introduction to $L_\infty$-algebras for classical field theories we shall formulate the model in terms of  $L_\infty$-algebra(s). Construction of an improved  DFT membrane sigma model is still work in progress, so we shall close this contribution with some insight obtained so far \cite{p3}.

\section{Courant sigma model}\label{sec:CSM}

The Courant sigma model was  constructed by examining consistent BV deformations of the abelian Chern-Simons gauge theory coupled with a  BF theory in Ref. \cite{ikeda}.  Related work in the context of open topological membranes can be found in Refs.\cite{Park1, Park2}. The model was also obtained in the framework of AKSZ construction of topological models satisfying the classical master equation \cite{AKSZ}.  This result is based on a theorem stating that there is a one-to-one correspondence between Courant algebroids and QP2-manifolds \cite{dee1}. Specializing  the general AKSZ construction to three worldvolume dimensions, this essentially means that given the data of a Courant algebroid one can construct a  unique three dimensional  sigma model, known as Courant sigma model  \cite{dee2}. 

 The classical action functional for this three-dimensional  topological field theory  is given as: 
\bea \label{csmif}
S[X,A,F]=\int_{\S_3}  \! \! F_i\w\dd X^i+\sfrac 12 \eta_{IJ}A^I\w\dd A^J-\rho^i{}_{I}(X)A^I\w F_i+\sfrac 16T_{IJK}(X)A^I\w A^J\w A^K~.
\eea
where  $i=1,\dots,d$ is a target space index, $ I=1,\dots,2d$ is the pull-back bundle index. The field content is made up of: scalar fields considered as components of maps $X=(X^i):\S_3\to {M}$, 1-forms $A\in \Omega^1(\S_3,X^{\ast}{E})$, and an auxiliary 2-form $F\in \Omega^2(\S_3,X^{\ast}T^{\ast}{M})$.  Fields $X^i$  are identified with the pull-backs of the coordinate functions, $X^i=X^*(x^i)$ on the target manifold $M$.
Here, $\eta$ is the $O(d,d)$-invariant constant metric:
\be \label{eta}
\eta=(\eta_{IJ})=\begin{pmatrix}
	0 & 1_{d} \\ 
	1_{d} & 0
 \end{pmatrix}~,
\ee
and $\rho^i{}_{{J}}(X)$  and $T_{IJK}(X)$   are functions corresponding to the anchor map and the  twist of the Courant algebroid.  The last term in the action \eqref{csmif} can be seen to generate a generalized Wess-Zumino term (WZ). 
For example, choosing $\rho^i{}_{j}=\delta^i{}_j, \;\r^{ij}=0$ and $T_{IJK}=H_{ijk},\;A^I=(q^i,p_j)$ produces:
\bea 
S[X,A,F]=\int_{\S_3}  \! \! F_i\w(\dd X^i-q^i)+q^i\w\dd p_i+\sfrac 16 H_{ijk}q^i\w q^j\w q^k~.\nn
\eea
On-shell, it is the topological sector of the string with a WZ-term.

Note that for a manifold with boundaries one can add both topological and non-topological terms \cite{CF1, Park1}:
\bea
S_b[X,A]=\int_{\partial\S_3}\!\!\sfrac 12 g_{IJ}A^I\w\ast A^J+\sfrac 12 {\cal B}_{IJ} A^I\w A^J~.\nn\eea
These boundary terms were used  in the description of T-dual backgrounds\footnote{Similarly, one can  use appropriate boundary conditions for on-shell fields \cite{Pavol2}.} in  Ref.\cite{p1}, 
 but in this contribution we shall assume a vanishing boundary.

The action \eqref{csmif} is invariant under
the  following infinitesimal gauge transformations \cite{ikeda} :
\bea\label{gt1}
&&\delta_\epsilon X^i=\r^i_J\epsilon^J~,\nn\\
&&\delta_\epsilon A^I=\dd\epsilon^I+\eta^{IN}T_{NJK}A^J\e^K-\eta^{IJ}\rho^i{}_Jt_i~, \\
&&\d_{\epsilon}F_m=-\dd t_m-\partial_m\r^j{}_J A^J\w t_j-\epsilon^J\partial_m\rho^{i}{}_{J}F_i+\sfrac 12\epsilon^J\partial_m T_{ILJ}A^I\w A^L~,\nn
\eea
where $\e^{I}$ and $t_{i}$ are scalar and 1-form gauge parameters, respectively, provided:
\begin{align}\label{gi}
\eta^{JK}\r^i{}_J(X)\r^j{}_K(X)&=0~,\nn\\
 2\r^m{}_{[K}(X)\partial_{\underline m}\rho^i{}_{J]}(X)-\r^i{}_N\eta^{NM}T_{MKJ}(X)&=0~,\\
3\eta^{MN} T_{M[JK}(X)T_{I]LN}(X)+3\rho^{m}{}_{[I}(X)\partial_m  T_{KJ]L}(X)+ \rho^{m}{}_{L}\partial_m T_{IJK}(X)&=0~.\nn
\end{align}
Furthermore, the closure of the algebra of gauge transformations gives:
\bea
[\delta_1,\delta_2]X^i&=&\r^i{}_{ J}\epsilon_{12}^{J}~, \; \epsilon_{12}^{ I}=\eta^{IJ}T_{JKL}\epsilon_1^{K}\epsilon_2^{L},\nn\\
\left[\delta_1,\delta_2\right]A^{I}&=&\delta_{12} A^{ I}- \eta^{IJ}\partial_mT_{JKL}\epsilon_1^{K}\epsilon_2^{ L} {\cal D}X^m~,\;
t_{12i}=\partial_iT_{KLJ}\,\epsilon_1^{K}\epsilon_2^{L}A^{J}+2\partial_i\r^j{}_{K}\epsilon^{K}_{[1}\,\,t_{2]j}~,\eea
where ${\cal D}X^m:=\dd X^i-\r^i{}_JA^J=0$ on $F$ equation of motion. 
We see that the gauge  transformations \eqref{gt1} define a first-stage reducible gauge symmetry, typical for gauge theories that include differential forms with degree larger than one \cite{Henneaux:1989jq,Gomis:1994he}. Moreover, the algebra of gauge transformations closes only on-shell. In the following we shall reformulate CSM with its rich gauge structure as an $L_\infty$-algebra.

\section{Courant sigma model as $L_\infty$-algebra}

The gauge symmetry of the Courant sigma model is based on an open algebra, i.e., on an algebra that closes only on-shell.  In order to properly define a CSM action functional, one should extend the classical action by additional fields (ghost etc.) into the BV-BRST framework \cite{ikeda, dee2}. The natural algebraic structure underlying the BV-BRST construction is in fact an $L_\infty$-algebra, see \cite{Barton, Jim,Jurco}.  In Ref.\cite{p2} we have constructed explicitly the $L_\infty$-algebra for BV CSM, but in this short note we restrict ourselves to classical theory.  We shall start this section with a short overview of $L_\infty$-algebras following Refs.\cite{Olaf, Jurco} and then exemplify the general construction on the Courant sigma model following \cite{p2}.

\subsection{$L_\infty$-algebra for classical field theories}

A ${L_\infty}${-algebra} or strong homotopy Lie algebra $(\mathsf{L},\mu_i)$ is a  graded vector space $\mathsf{L}$  with a collection of  graded totally antisymmetric  multilinear maps:
\be
\mu_i:\underbrace{\mathsf{L}\times\cdots\times \mathsf{L}}_{\text{$i$-times}}\to \mathsf{L}~. \nonumber
\ee
of degree $2-i$ which satisfy the homotopy Jacobi identities:
\bea\label{hom}
\sum_{j+k=i}\sum_\sigma\chi(\sigma;l_1,\ldots,l_i)(-1)^{k}\mu_{k+1}(\mu_j(l_{\sigma(1)},\ldots,l_{\sigma(j)}),l_{\sigma(j+1)},\ldots,l_{\sigma(i)})&=0~;\\
\forall i\in\mathbb{N}\quad \forall l_i&\in\mathsf{L}\nonumber~.
\eea
where	$\chi(\sigma;l_1,\ldots,l_i)$ is the graded Koszul sign that includes the sign from the parity of the permutation $\sigma$  ordered as: $\sigma(1)<\cdots<\sigma(j)$ and $\sigma(j+1)<\cdots<\sigma(i)$.
Let us first examine a few relations resulting from the general expression \eqref{hom}.  Taking the free summation index $i=1$ we get $\mu_1^2(l)=0$, i.e., the  map $\mu_1$ is differential on $\mathsf{L}$. For 
$i=2$ we have:
\bea
\mu_1(\mu_2(l_1,l_2)=\mu_2(\mu_1(l_1),l_2)+(-1)^{1+|l_1||l_2|}\mu_2(\mu_1(l_2),l_1)~, \eea
where $|l_i|$ is the $L_\infty$-degree of $l_i$, specifying that  $\mu_1$ is a derivation of (graded) bracket $\mu_2$.  Finally, for ${ i=3}$  we get the generalized Jacobi identity for bracket $\mu_2$ controlled by  $\mu_3$:
\begin{align}
\mu_1(\mu_3(l_1,l_2,l_3))&=\mu_2(\mu_2(l_1,l_2),l_3)-(-1)^{|l_2||l_3|}\mu_2(\mu_2(l_1,l_3),l_2)+{}\nn\\
&\phantom{\,=\,}+(-1)^{|l_1|(|l_2|+|l_3|)}\mu_2(\mu_2(l_2,l_3),l_1)-\mu_3(\mu_1(l_1),l_2,l_3)+{}\\
&\phantom{\,=\,}+(-1)^{|l_1||l_2|}\mu_3(\mu_1(l_2),l_1,l_3)-(-1)^{|l_3|(|l_1|+|l_2|)}\mu_3(\mu_1(l_3),l_1,l_2)~,\nn
\end{align}
Standard examples include a Lie algebra with $\mu_2$ being the Lie bracket and all other maps vanishing, and a differential graded Lie algebra obtained for $\mu_i=0$ for $i\geq3$.

Elements of graded vector space $\mathsf{L}=\bigoplus_i\mathsf{L}_i$  form a cochain complex as $\mu_1$ is a differential: 
\bea
\cdots   \overset{\mu_1}{\longrightarrow}   \mathsf{L}_{i} \overset{\mu_1}{\longrightarrow}   \mathsf{L}_{i+1} \overset{\mu_1}{\longrightarrow}     \cdots \nn
\eea
 In order to describe classical field theory we associate with each subspace a particular field content, including fields, gauge parameters, equations of motion and Noether identities.
\bea
\begin{array}{ccccccccccccc}
\cdots & \overset{\mu_1}{\rightarrow} & \mathsf{L}_{-1}&\overset{\mu_1}{\rightarrow}  & \mathsf{L}_0&\overset{\mu_1}{\rightarrow}  & \mathsf{L}_1 &\overset{\mu_1}{\to} & \mathsf{L}_2 &\overset{\mu_1}{\to} & \mathsf{L}_3&\overset{\mu_1}{\rightarrow}  & \cdots \\
& &{\scriptstyle{\rm h.\;gauge}}& &{\scriptstyle{\rm gauge}}&  &{\scriptstyle{\rm physical}}&  &{\scriptstyle{\rm eoms}} &  &{\scriptstyle{\rm Noether}} &&\\[-2mm]
& &{\scriptstyle{\rm parameters}}& &{\scriptstyle{\rm parameters}}&  &{\scriptstyle{ \rm fields}}&  &{\scriptstyle{ }} &  &{\scriptstyle{\rm identities}}&&\\[-2mm]
\end{array}\nn
\eea
Take $a\in \mathsf{L}_1$ to be the gauge potential. The corresponding curvature is:
\be
f\equiv\mu_1(a)+\frac 1 2 \mu_2(a,a)+\cdots=\sum_{i\geqslant1}\frac{1}{i!}\mu_i(a,\ldots,a)~.\nn
	\ee
If $f=0$   we obtain the generalized  Maurer-Cartan equation. 
Next, for  $c_0\in \mathsf{L}_0$  being the level zero gauge parameter, we can write gauge transformations as
\bea\label{gtg}
\delta_{c_0}a&=&\sum_{i\geqslant0}\frac{1}{i!}\mu_{i+1}(a,\ldots,a,c_0)~,\nn\\
\delta_{c_0}f&=&\sum_{i\geqslant0}\frac{1}{i!}\mu_{i+2}(a,\ldots,a,f,c_0)~.
\eea
If a theory contains higher gauge symmetries we will have higher ({level $k$) gauge parameters $c_{-k}\in\mathsf{L}_{-k}$, $k>0$, with infinitesimal gauge transformations given by:
\[
\delta_{c_{-k-1}}c_{-k}=\sum_{i\geqslant0}\frac{1}{i!}\mu_{i+1}(a,\ldots,a,c_{-k-1})~.
\]
The algebra of gauge transformations:
\be
[\delta_{c_0},\delta_{c'_0}]a=\delta_{c''_0}a+\sum_{i\geqslant0}\frac{1}{i!}(-1)^i\mu_{i+3}(f,a,\ldots,a,c_0,c'_0)~,\nn
\ee
with:
$$c''_0=\sum_{i\geqslant0}\frac{1}{i!}\mu_{i+2}(a,\ldots,a,c_0,c'_0)~.$$
 closes up to $f=0$, which can be realized either as a constraint or as an equation of motion, the latter resulting in an open algebra.  However, to implement $f=0$ as a variational equation of motion resulting from an action functional, we need additional input -- a cyclic pairing satisfying a certain compatibility condition. In that case we define a cyclic   ${L_\infty}${-algebra} as an $L_\infty$-algebra  endowed with a graded symmetric  non-degenerate bilinear pairing 
$	\langle \,\cdot\,, \,\cdot\, \rangle_{\mathsf{L}}:\mathsf{L}\times\mathsf{L}\to\mathbb{R}$
	that satisfies the cyclicity condition:
	\begin{align}
	\langle l_1,\mu_i(l_2,\ldots,l_{i+1})\rangle_{\mathsf{L}}=(-1)^{i+i(|l_1|+|l_{n+1}|)|l_{i+1}|\sum_{j=1}^{i}|l_j|}\langle l_{i+1},\mu_i(l_1,\ldots,l_i)\rangle_{\mathsf{L}}~.\nn
	\end{align}
Then one can define an action whose stationary point is the Maurer-Cartan equation:
	\begin{align}\label{MC}
	S_{\mathrm{MC}}[a]\equiv\sum_{i\geqslant1}\frac{1}{(i+1)!}\langle a,\mu_i(a,\ldots,a)\rangle_{\mathsf{L}}~.
	\end{align}

As a conclusion of  this short summary of ${L_\infty}$-structures we introduce an important class of ${L_\infty}${-algebras}  induced by  the tensor product of an $L_\infty$-algebra  with a differential graded commutative algebra.  In the following sections  we shall use an $L_\infty$-algebra obtained as a tensor product of the de Rham complex on a manifold $M$, $(\Omega^\bullet(M),\dd)$     with an ${L_\infty}${-algebra}:
\bea
\mathsf{L'}\equiv\Omega^\bullet(M,\mathsf{L})\equiv\bigoplus_{k\in\mathbb{Z}}\Omega^\bullet_k(M,\mathsf{L}),\quad \Omega^\bullet_k(M,\mathsf{L})\equiv\bigoplus_{i+j=k}\Omega^i(M)\otimes\mathsf{L}_j~,\nn
\eea
where the  higher products are defined as:
\bea\label{tensor}
\mu'_1(\alpha_1\otimes l_1)&=&\dd \alpha_1\otimes l_1+(-1)^{|\alpha_1|}\alpha_1\otimes\mu_1(l_1)~,\label{eq:muprime1}\nn\\
\mu'_i(\alpha_1\otimes l_1,\ldots,\alpha_i\otimes l_i)&=&
(-1)^{i\sum_{j=1}^{i}|\alpha_j|+\sum_{j=0}^{i-2}|\alpha_{i-j}|\sum_{k=1}^{i-j-1}|l_k|}(\alpha_1\wedge\cdots \wedge\alpha_i)\otimes\mu_i(l_1,\ldots,l_i)~,\label{eq:muprimei}\nn\\
&& \qquad\qquad\qquad\!\!\forall i\geqslant 2,\quad \alpha_1,\ldots,\alpha_i\in\Omega^\bullet(M),\quad l_1,\ldots,l_i\in\mathsf{L}~.
\eea
This tensored algebra  $(\mathsf L', \mu'_i)$   is cyclic provided  $(\mathsf{L},\mu)$ is cyclic and $M$ is an oriented, compact cycle.

\subsection{$L_\infty$-algebra for Courant sigma model}

In the following we shall apply the above formalism to cast the Courant sigma model into an ${L_\infty}$ structure. In  action \eqref{csmif} we first identify $\(X,A,F\)$ as physical fields, then perturbatively expand structure functions $\r(X)$  and $T(X)$   around $X=0$, resulting in  an action with an infinite number  of interaction terms:\footnote{We introduce the shorthand $f(0)\equiv f$ and $\partial f\big|_0\equiv\partial f$ for any function $f$ of $X$ evaluated at 0, and when the full function is meant the argument will be explicitly written.}
  \begin{align}\label{eq:csmactiontaylor}
 S[X,A,F]=\int_{\S_3}  \! \! &F_i\w\dd X^i+\sfrac 12 \eta_{IJ}A^I\w\dd A^J-\rho^i{}_{I}A^I\w F_i+\sfrac 16T_{IJK}A^I \w A^J\w A^K{}-\nn\\
  \phantom{\int_{\S_3}  \! \! }&-X^{i_1}\partial_{i_1}\rho^i{}_{I}A^I\w F_i-\cdots-\sfrac{1}{n!}X^{i_1}\cdots X^{i_n}\partial_{i_1}\cdots\partial_{i_n}\rho^i{}_{I}A^I\w F_i-\cdots+{}\nonumber\\
 \phantom{\int_{\S_3}  \! \! }&+\sfrac 1 {12}X^{i_1}X^{i_2}\partial_{i_1}\partial_{i_2}T_{IJK}A^I \w A^J \w A^K+\cdots+{}\nn\\
 \phantom{\int_{\S_3}  \! \! }&+\sfrac 1 {6\cdot n!} X^{i_1}\cdots X^{i_n}\partial_{i_1}\cdots\partial_{i_n}T_{IJK}A^I \w A^J \w A^K+\cdots. 
 \end{align}
 In Sect. \ref{sec:CSM} we have   seen that the gauge symmetry is mediated by two gauge parameters, one scalar $\epsilon$ and one 1-form $t$.  Although not exemplified at the classical level,  we also have a higher gauge parameter $v$ of level 1, which mediates the transformation of level zero parameters. Therefore, the complete (classical) field content is:
\be
\begin{split}
a=X+A+F&\in\Omega^0(M,\mathsf{L}_1)\oplus\Omega^1(M,\mathsf{L}_0)\oplus\Omega^2(M,\mathsf{L}_{-1})~,\nn\\
c_0=\epsilon+ t&\in \Omega^0(M,\mathsf{L}_0)\oplus\Omega^1(M,\mathsf{L}_{-1})~,\nn\\
c_{-1}=v&\in\Omega^0(M,\mathsf{L}_{-1})~,\nn
\end{split}
\ee
Introducing notation ${\cal D}E$ for the equation of motion of physical field $E=\(X,A,F\)$ we can represent the content of graded vector space $\mathsf{L}$ in the following table.
\[
\begin{array}{cccccccccc}
 &\mathsf{L'}_{-1}&\overset{\mu'_1}{\longrightarrow}  & \mathsf{L'}_0&\overset{\mu'_1}{\longrightarrow}  & \mathsf{L'}_1 &\overset{\mu'_1}{\longrightarrow} & \mathsf{L'}_2 &\overset{\mu'_1}{\longrightarrow} & \mathsf{L'}_3 \\
&{\scriptstyle{\rm h.\;gauge}}& &{\scriptstyle{\rm gauge}}&  &{\scriptstyle{\rm physical}}&  &{\scriptstyle{\rm equations}} &  &{\scriptstyle{\rm Noether}} \\[-2mm]
&{\scriptstyle{\rm parameters}}& &{\scriptstyle{\rm parameters}}&  &{\scriptstyle{ \rm fields}}&  &{\scriptstyle{\rm of\;motion}} &  &{\scriptstyle{\rm identities}}\\[-2mm]
   &  & & &  & & & & & \\
  \mathsf{L}_{-1} &v_i & & t_i & &  F_i &  & \mathcal{D}F_i & &  \\
 \hspace{-5mm}{\scriptstyle\mu_1} \downarrow   &  & & & & &  \\
 \mathsf{L}_0& &  & \epsilon^I &  & A^I & & \mathcal{D}A^I & & \\
\hspace{-5mm}{\scriptstyle\mu_1}\downarrow & & & & & & & & \\
 \mathsf{L}_1 &  & & &  & X^i &  &  \mathcal{D}X^i & & 
\end{array}\]
Once we have placed the fields in their appropriate homogeneous subspaces, we must define all the products. Our selection for the nonvanishing higher products of $\mathsf{L}$ is:
\begin{align}
\mathsf{L}_1&\ni & \mu_n(l_{(1)1},\ldots,l_{(1)n-1},l_{(0)})&=l^{i_1}_{(1)1}\cdots l^{i_{n-1}}_{(1)n-1}\partial_{i_1}\cdots\partial_{i_{n-1}}\rho^i{}_Il^I_{(0)}~,\label{eq:mu10}\\
\mathsf{L}_0&\ni & \mu_n(l_{(1)1},\ldots,l_{(1)n-1},l_{(-1)})&=-l^{i_1}_{(1)1}\cdots l^{i_{n-1}}_{(1)n-1}\partial_{i_1}\cdots\partial_{i_{n-1}}\rho^i{}_Jl_{(-1)i}\eta^{IJ}~,\nn\\
\mathsf{L}_{-1}&\ni & \mu_m(l_{(1)1},\ldots,l_{(1)m-2},l_{(-1)},l_{(0)})&=-l^{i_1}_{(1)1}\cdots l^{i_{m-2}}_{(1)m-2}\partial_{i_1}\cdots\partial_{i_{m-2}}\partial_i\rho^j{}_Il_{(-1)j}l^I_{(0)}~,\nn\\
\mathsf{L}_0&\ni & \mu_m(l_{(1)1},\ldots,l_{(1)m-2},l_{(0)1},l_{(0)2})&=l^{i_1}_{(1)1}\cdots l^{i_{m-2}}_{(1)m-2}\partial_{i_1}\cdots\partial_{i_{m-2}}T_{JKL}l_{(0)1}^Kl_{(0)2}^L \eta^{IJ}~,\nn\\
\mathsf{L}_{-1}&\ni & \mu_r(l_{(1)1},\ldots,l_{(1)r-3},l_{(0)1},l_{(0)2},l_{(0)3})&=l^{i_1}_{(1)1}\cdots l^{i_{r-3}}_{(1)r-3}\partial_{i_1}\cdots\partial_{i_{r-3}}\partial_iT_{IJK}l_{(0)1}^I l_{(0)2}^Jl_{(0)3}^K~,\nn
\end{align}
where $n\geqslant1$, $m\geqslant2$ and $r\geqslant3$, and $l_{(i)}\in\mathsf{L}_i$. We defined the products by comparing the general expression for the MC action \eqref{MC} with \eqref{eq:csmactiontaylor}, but they can also be obtained from the (expanded) homological vector $Q$ ~\cite{AKSZ,Voronov} defined for a Courant algebroid described as  a QP2-manifold in \cite{dee2}. The products for tensored algebra  $(\mathsf L', \mu'_i)$  can be easily deduced from the general expression \eqref{tensor} and the above expressions for $(\mathsf L, \mu_i)$ . Using  expressions \eqref{eq:mu10} for products one can recast the gauge transformation \eqref{gt1} of the Courant sigma model in general form \eqref{gtg}. 

Next, we  define a consistent inner product:
\bea\label{inner}
 \langle l_{(0)1}, l_{(0)2}\rangle\equiv\eta_{IJ}l_{(0)1}^I l_{(0)2}^J,  \langle l_{(1)}, l_{(-1)}\rangle&\equiv l_{(1)}^i l_{(-1)i}, & \langle l_{(-1)}, l_{(1)}\rangle\equiv -l_{(-1)}^i l_{(1)i}~,
 \eea
resulting in the MC action:
\begin{align}\label{mcc}
 S_{\mathrm{MC}}[X,A,F]&=\langle \dd X,F\rangle+\sfrac 1 2 \langle A,\dd A\rangle+\sum_{n\geqslant 0}\sfrac{1}{n!}\langle A,\mu_{n+1}(X,\ldots,X,F)\rangle+{}\nn\\
 &\phantom{\,=\,} +\sfrac 1 6 \sum_{n\geqslant0}\sfrac{1}{n!}\langle A,\mu_{n+2}(X,\ldots,X,A,A)\rangle~.
 \end{align}
Note that the physical field $A$  is a section of the pull-back bundle  $X^*E$ where there is no natural structure defined; in particular the bracket of sections  is
not the Courant bracket. Thus one can think of $L_\infty$-products as defining relations for the relevant structures on sections of the pull-back bundle. Likewise, the inner product \eqref{inner} defines a pairing on the pull-back bundle $X^*E$. 

The equations of motions are obtained directly from the action \eqref{mcc}:
 \begin{align*}
 f_{1}&=\dd X-\sum_{n\geqslant0}\sfrac{1}{n!}\mu_{n+1}(X,\ldots,X,A)~,\\
 f_0&=\dd A+\sum_{n\geqslant0}\sfrac{1}{n!}\mu_{n+1}(X,\ldots,X,F)+\sfrac 1 2\sum_{n\geqslant0}\sfrac{1}{n!}\mu_{n+2}(X,\ldots,X,A,A)~,\\
 f_{-1}&=\dd F-\sum_{n\geqslant0}\sfrac{1}{n!}\mu_{n+2}(X,\ldots,X,F,A)-\sfrac{1}{3!}\sum_{n\geqslant0}\sfrac{1}{n!}\mu_{n+3}(X,\ldots,X,A,A,A)~.
 \end{align*}
 With the aid of \eqref{eq:mu10}  it becomes obvious these correspond to the equations of motion of action \eqref{csmif} or \eqref{eq:csmactiontaylor}:
 \begin{align}
 \mathcal{D}X^i&=\dd X^i-\rho^i{}_J(X)A^J~,\nn\\
 \mathcal{D}A^I&=\dd A^I-\eta^{IJ}\rho^j{}_J(X)F_j+\sfrac 1 2 \eta^{IJ}T_{JKL}(X)A^K\w A^L~,\\
 \mathcal{D}F_i&=\dd F_i+\partial_i\rho^j{}_J(X)A^J\w F_j-\sfrac{1}{3!}\partial_iT_{IJK}(X)A^I\w A^J\w  A^K\label{eq:Xeom}.~\nn
 \end{align}

Finally, we need to check homotopy Jacobi identities \eqref{hom} of products \eqref{eq:mu10}. As explicitly shown in \cite{p2}, there exist 7 non-trivial identities of which 3 are independent. These three sets of
homotopy conditions for the higher products are actually all terms in the Taylor expansions of classical gauge invariance conditions \eqref{gi} and they belong, as expected,  to the space $\mathsf{L'}_3$ of Noether identities.

\section{On the relation between CA and DFT}

As already mentioned in the introduction, the DFT symmetry algebra reduces to the one of a CA on the solutions of strong constraint.
However, to understand the geometric origin of DFT data and strong constraint itself, one has to go  a step further.  Starting with the observation  that in DFT one doubles the target space, while a CA is defined on a doubled bundle,  in Ref.\cite{p1}  it was proposed that one should  construct a `large'  CA defined over $2d$ dimensional manifold and then recover DFT data by suitable projection.   
One starts from a CA structure defined on  a   bundle $\mathbb{E}=(T\oplus T^{\ast}){\cal M}$  over the doubled configuration space ${\cal M}$ with sections
  $\A\in\mathbb{E}$ 
\be
\A:=\A_{V}+\A_{F}=\A^I\partial_I+\widetilde \A_I\dd\X^I,\; I=1,\ldots, 2d~.\nn\ee 
Introducing an invariant $O(d,d)$ metric\footnote{This metric should not be confused with $O(2d,2d)$ invariant metric on the large Courant algebroid.} one can define the following splitting: 
$\mathbb{E}=(T\oplus T^{\ast}){\cal M}=L_+\oplus L_-$;
\be
\A=\A_+^Ie_I^++\A_-^Ie_I^-~, \quad \text{where} \quad e^{\pm}_I=\partial_I\pm \eta_{IJ}\dd\X^J~.\nn\ee
Using the projection:
\bea 
p: \mathbb{E} \to L_+~; \quad(\A_V,\A_F) \mapsto \A_+:=A~,\nn
\eea 
one can systematically project the bracket, bilinear and structure functions (anchor map and twist) of the large CA to obtain DFT vectors, the C-bracket and generalized Lie derivative with properties defining a DFT algebroid. 
Using the data of this DFT algebroid a membrane sigma model was proposed:
 \be S[\X , A,F]=\int_{\Sigma_3}\left(F_I\w \dd\X^I+\eta_{IJ}A^I\w\dd A^J-\widehat \rho^{I}{}_{J}(\X)A^J\w F_I+\sfrac 13 \widehat T_{IJK}(\X)A^I\w A^J\w A^K\right)~,\nn\ee
 where  one introduced  maps $\X=(\X^I):\S_3\to {\cal M}$, 1-form $A\in \Omega^1(\S_3,\X^{\ast}L_+)$, and an auxiliary 2-form $F\in \Omega^2(\S_3,\X^{\ast}T^{\ast}{\cal M})$. The structure functions $\widehat\r$ and $\widehat T$ obtained through the projection  were then related to DFT fluxes and their Bianchi identities using the flux formulation of DFT \cite{dftflux3}. This was realized  by parametrizing  the projected anchor map $\r_+$  in terms of the generalized bein components of DFT. The identification imposes:\footnote{Additionally, it implies the helpful identity $\eta_{KL}\widehat\rho^K{}_I\widehat\rho^L{}_J=\eta_{IJ}$,
heavily used in calculations.}
\bea
\eta^{IJ}\widehat\rho^K{}_I\widehat\rho^L{}_J=\eta^{KL}~,
\eea
and reveals the source of the strong constraint in DFT   as coming from  constraining the generalized bein of DFT to be an element of $O(d,d)$.   Further consistency conditions defining the DFT algebroid:
\begin{align} \label{dft}
 2\widehat\rho^{L}{}_{[I}\partial_{\underline{L}}\widehat\rho^{K}{}_{J]}-\widehat\rho^{K}{}_{M}\eta^{MN}\widehat T_{NIJ}&=\eta_{LL'}{\widehat\rho^{L}{}_{[I}\partial^K\widehat\rho^{L'}{}_{J]}}~, \nn  \\
3\eta^{MN} \widehat T_{M[IJ}\widehat T_{KL]N}-4\widehat\rho^{M}{}_{[I}\partial_{\underline M} \widehat T_{JKL]}&=3\eta_{MM'}\eta_{NN'}\eta^{PP'}\widehat\rho^M{}_{[I}\partial_{\underline P}\widehat\rho^{M'}{}_J\,\widehat\rho^N{}_{K}\partial_{\underline P'}\widehat\rho^{N'}{}_{L]}~,
\end{align}
were used to demonstrate that one should use the strong constraint to guarantee the gauge invariance of the DFT sigma model and  (on-shell) closure of the gauge algebra.

The constructed membrane sigma model   allows for the description of  a standard T-duality orbit of geometric and non-geometric flux backgrounds  in a unified way \cite{p1}. However, the necessity to enforce strong constraint limits its physical applications.  As  mentioned in the introduction, solving the strong constraint effectively reduces physical field dependence to only  half of the coordinates of doubled space. For example, when the compactification scale is much bigger than the string size, the winding modes are heavy and at low energies  this corresponds to the usual situation where there is no dependence on dual coordinates. Oppositely, in the T-dual description, if the compactification scale is small, then the momentum modes are heavy, and DFT only depends on dual coordinates. However, in order to describe the process of decompactification one should have a consistent description that includes both momentum and winding modes/coordinates. And precisely this process of decompactification is at heart of the string-gas cosmology proposal \cite{BV,BR}.  
Furthermore, recent developments in string theory have suggested that the low-energy effective dynamics of closed strings in non-geometric flux compactifications may be governed by noncommutative and
even nonassociative deformations of gravity \cite{BlumenhagenNA1, BlumenhagenNA2, Mylonas, BlumenhagenNA3,Mylonas2}. However these backgrounds are, in general, not solutions of the strong constraint \cite{BlumenhagenNA3,p1}.  Finally, in the approach of the generalized Scherk-Schwarz compactification in  DFT it was shown that one could consistently relax  this constraint \cite{mariana,dftflux3}. 

Here we want to argue that recasting the problem  into an $L_\infty$-framework could indicate a way of  relaxing  this strong constraint. The first observation we make  is that the relevant structure for the case of a DFT algebroid is that of curved $L_\infty$-algebras. These are $L_\infty$-algebras that include a $\mu_0$ map that corresponds to a constant element\footnote{From the point of view of QP manifolds this constant comes from the expansion of  homological vector Q  outside of the fixed point.} of the homogeneous subspace of degree 2 in our conventions. The homotopy relations are again written in similar form as \eqref{hom}, however, now every expression of a fixed order in $i$ has an additional term coming from $\mu_0$. In particular:
$$i=1:\;\;\mu_1^2(l)= \mu_2(\mu_0,l)~,$$
$$i=2: \;\;\mu_1(\mu_2(l_1,l_2)-\mu_2(\mu_1(l_1),l_2)-(-1)^{1+|l_1||l_2|}\mu_2(\mu_1(l_2),l_1)=-\mu_3(\mu_0,l_1,l_2)~,$$
Importantly, map $\mu_1$ is no longer a differential  and there is no (co)chain complex underlying a curved $L_\infty$-algebra.

In DFT framework the strong condition is expressed as a differential equation:
$$\eta^{IJ}\partial_I\partial_J (\ldots)=0~,$$
where derivatives act on products of fields.  This lead us to propose a curved $L_\infty$- algebra of DFT defined on the following graded space.
\[
\begin{array}{ccccc}
\mathsf{L}_{-1}&\oplus  & \mathsf{L}_0 &\oplus &\mathsf{L}_2 \\
f\in C^\infty({\cal M}) & & e\in \Gamma(L_+) & & \mu_0
\end{array}\]
Following the construction of the $L_\infty$- algebra for a CA \cite{rw} one can define a curved $L_\infty$- algebra for the DFT algebroid \cite{p3} where:
$$\mu_0=\eta^{IJ}\partial_I\otimes\partial_J~,$$
is a constant symmetric 2-vector, and homotopy relations reproduce the defining relations of DFT algebroid \eqref{dft}. 
The physical relevance of constant map $\mu_0$ in the context of DFT is still under investigation. Note, however, that in Ref. \cite{Barton}, Zwiebach identified $L_\infty$-algebras as the structure  underlying  closed string field theory and related the $\mu_0$ map to the backgrounds which are not conformal. In DFT context, this nicely fits with observation that on one hand the non-associative backgrounds violate the strong constraint of DFT, and that on the other hand they are incompatible with on-shell conformal symmetry of string theory, see discussion in Ref.\cite{RR}.  To further the  understanding of the constraint structure of DFT we propose to use the homotopy relations of $L_\infty$-algebra  to bootstrap the gauge invariant membrane sigma model of relevance to the unconstrained theory. We plan to report on  progress of this direction in forthcoming work  \cite{p3}.

\acknowledgments
We thank the organizers of Humboldt Kolleg Frontiers in Physics:
From the Electroweak to the Planck Scales and  the Alexander von Humboldt Foundation. 
The work was supported by the Croatian Science Foundation under the Project IP-2019-04-4168
and  by the European Union through the European Regional Development Fund - the Competitiveness and Cohesion Operational Programme (KK.01.1.1.06).

\end{document}